%% file: main.tex
\documentclass[sigconf,review=false]{acmart}
\AtBeginDocument{%
  \providecommand\BibTeX{{%
    \normalfont B\kern-0.5em{\scshape i\kern-0.25em b}\kern-0.8em\TeX}}}

\copyrightyear{2019}
\acmYear{2019}
\setcopyright{rightsretained}

\acmConference{LAK 2020}{March 23 - March 27, 2020}{Frankfurt, Germany}
\acmDOI{10.1145/3306307.3328180}
\acmISBN{978-1-4503-6317-4/19/07}
\acmBooktitle{LAK 2020}

\newcommand{\yong}[1]{\textcolor{blue}{\textit{[Yong: #1]}}}

\usepackage{subcaption}
\usepackage{amsmath}
\usepackage{multirow}

\begin{document}

% \title{Predicting Students' Performance in Interactive Math Question Pools Using Mouse Movement Sequences}
% \title{Predicting Students' Performance in Interactive Online Question Pools Using Mouse Movement Sequences}
% \title{Predicting Students' Performance in Interactive Online Question Pools Using Interaction Features}
\title{Predicting Student Performance in Interactive Online Question Pools Using Mouse Interaction Features}

% \author{Huan Wei}
% % \email{trovato@corporation.com}
% % \orcid{1234-5678-9012}
% % \author{G.K.M. Tobin}
% \affiliation{%
%   \institution{Hong Kong University of Science and Technology}
%   \city{HongKong SAR}
%   \state{China}
% %   \postcode{43017-6221}
% }

\author{Huan Wei, Haotian Li, Meng Xia, Yong Wang, Huamin Qu}
\affiliation{%
  \institution{Department of Computer Science and Engineering, HKUST, Hong Kong SAR, China}
}
\email{{hweiad, hlibg, iris.xia, ywangct}@connect.ust.hk;{huamin}@cse.ust.hk}

% \affiliation{%



\renewcommand{\shortauthors}{Huan Wei et al.}

\begin{abstract}
Modeling student learning and further predicting the performance is a well-established task in online learning and is crucial to personalized education by recommending different learning resources to different students based on their needs. Interactive online question pools (e.g., educational game platforms), an important component of online education, have become increasingly popular in recent years. However, most existing work on student performance prediction targets at online learning platforms with a well-structured curriculum, predefined question order and accurate knowledge tags provided by domain experts. It remains unclear how to conduct student performance prediction in interactive online question pools without such well-organized question orders or knowledge tags by experts. In this paper, we propose a novel approach to boost student performance prediction in interactive online question pools by further considering student interaction features and the similarity between questions. Specifically, we introduce new features (e.g., think time, first attempt, and first drag-and-drop) based on student mouse movement trajectories to delineate students' problem-solving details. In addition, heterogeneous information network is applied to integrating students' historical problem-solving information on similar questions, enhancing student performance predictions on a new question. We evaluate the proposed approach on the dataset from a real-world interactive question pool using four typical machine learning models. The result shows that our approach can achieve a much higher accuracy for student performance prediction in interactive online question pools than the traditional way of only using the statistical features (e.g., students' historical question scores) in various models. We further discuss the performance consistency of our approach across different prediction models and question classes, as well as the importance of the proposed interaction features in detail.
\end{abstract}


% \ccsdesc[500]{Applied computing~Collaborative learning}



\maketitle
\input{src/01_introduction.tex}

\input{src/02_relatedwork.tex}
\input{src/03_background.tex}
\input{src/04_method.tex}

\input{src/05_evaluation.tex}

\input{src/06_discussion.tex}
\input{src/07_conclusion.tex}

\begin{acks}
The work is funded by HK ITF under grant ITS/388/17FP. Y. Wang is the corresponding author. Thank Trumptech (Hong Kong) Limited for the data support and Min Xu for proofreading.
\end{acks}

\bibliographystyle{ACM-Reference-Format}
\bibliography{ref}

\end{document}

%% file: src/01_introduction.tex
\section{Introduction}

% 1. 1) The importance of performance prediction in online question pools
%    a) need to estimate the student's potential performance to provide students with questions of appropriate difficulty level
%    b) dropout prediction
%    c) Knowledge tracing --> performance prediction --> intelligent curriculum design 
%    2) Introduce interactive online question pools
With the rapid development of digital technologies, online education has become increasingly popular in the past decade, which provides students with easy access to various online learning materials such as massive open online courses (MOOCs) and different online question pools.
The availability of these online learning materials and student activity logs has made it possible to model student learning and further predict their performance~\cite{kaser2017modeling}. It has become a well-established task in computer-aided education, since student performance prediction can help course designers to better handle the possible dropout, improve the retention rate of online learning platforms~\cite{manrique2019analysis}, and provide students with personalized education to enhance student learning by recommending different learning resources to them based on their different needs~\cite{piech2015deep,yeung2018addressing,thaker2019comprehension,ding2018effective}.
% \yong{My task: need to refer to Figure 1 in some part.}

% With the rapid development of online education,
Online question pools, consisting of a collection of questions, have become increasingly popular for helping students to practice their knowledge and skills~\footnote{\href{https://help.blackboard.com/Learn/Instructor/Tests_Pools_Surveys/Orig_Reuse_Questions/Question_Pools}{https://help.blackboard.com/}}. For example, there have been many popular online question pools that provide students hands-on exercises to practice their programming skills (e.g., LeetCode\footnote{\href{https://leetcode.com/}{\url{https://leetcode.com/}}}, Topcoder\footnote{\href{https://www.topcoder.com/}{\url{https://www.topcoder.com/}}}) and mathematics skills (e.g., Learnlex\footnote{\href{https://mad.learnlex.com}{\url{https://mad.learnlex.com/login}}}, Math Playground\footnote{\href{https://www.mathplayground.com/}{https://www.mathplayground.com/}}).
% In addition, many questions in the online question pools have been designed to involve user interactions to further %make the whole learning process more interesting, 
% make it a fun way to practice important skills~\cite{onlineInteractiveGames2019}.
% Figure~\ref{example} shows an example of such interactive online questions, where students are asked to interactively use mouse to move the red dot to satisfy certain requirements. \yong{To be done.}
The \textit{interactive online question pool}, as one kind of the online question pools, comprises interactive questions where interactions such as mouse movement and drag-and-drop are often needed. For example, some online educational games have been designed to involve user interactions to make it a fun way to practice important skills~\cite{irene2010multimedia}~\footnote{\href{https://www.weareteachers.com/the-best-online-interactive-math-games-for-every-grade-level/}{https://www.weareteachers.com/}}.
% , since the questions are interesting and working on them is like playing games.
Despite the popularity and importance of interactive online question pools,
little work has been done to model student learning and predict their performance in online question pools.

Student performance prediction has been widely explored in education community and is a critical step for downstream tasks, such as recommending an adaptive learning path~\cite{xia2019peerlens} or assisting students at an early stage~\cite{chen2016dropoutseer}. 
% Extensive studies have been conducted on student learning modeling and student performance prediction.
However, most of them focus on student performance prediction on the MOOCs platforms (e.g, Coursera, EdX, Khan Academy),
and little work has been done on interactive online question pools.
Compared with student performance prediction on MOOCs platforms, it is more challenging to predict student performance on interactive online question pools due to two major reasons.
First, there is \textbf{\textit{no predefined question order}} that users can follow; students need to explore the questions by themselves when working on the interactive online question pools.
Moreover, for some interactive online question pools, their questions only have \textbf{\textit{rough knowledge tags}} annotated by domain experts, which are not accurate enough to evaluate the similarity among questions.
% different from other learning platforms where questions are usually labelled with knowledge tags by experts to denote the knowledge concept, it is often deliberately omitted in online question pools to avoid revealing the real intent of the questions (\textbf{\textit{no knowledge tags}}).
% Moreover, different from other learning platforms where \textbf{\textit{knowledge tags}} are usually labeled for each question by experts to denote the knowledge concept. It is often deliberately avoided to reveal the real intent of the questions in these online question pools for examination purpose~\cite{xia2019peerlens}.
As will be introduced in Section~\ref{sec_performance_prediction}, the major models for student performance prediction include Bayesian knowledge tracing ~\cite{corbett1994knowledge,pardos2011kt}, deep knowledge tracing ~\cite{piech2015deep,yeung2018addressing} and traditional machine learning models ~\cite{chen2016dropoutseer, manrique2019analysis,o2018student}.
% \yong{Huan, pls add references here.}
% (e.g., regression models and support vector machine (SVM)). 
However, almost all of these models intrinsically depend on the course curriculum or the predefined question order and Bayesian knowledge tracing models also require knowledge tags, making them inapplicable to student performance prediction on interactive online question pools.

To handle the above challenges, we propose a novel method by introducing a set of new features based on interactions between students and questions to perform student performance prediction in interactive online question pools. 
% Specifically, we focus on introducing new features.
% that have been researched in education and psychology fields to predict student performance by further deliberately considering student mouse movement interactions and the similarity between different questions.
% \yong{Huan, pls double check if the claims below are correct or not. checked}
% Existing studies usually use student profiles (e.g, the age, gender, year of study) and their past problem-solving information (e.g., the accuracy and time cost) to model student learning and their performance.
Specifically, we utilize the mouse  movement trajectories, which consists of the mouse interaction timestamp,  mouse event type and mouse coordinates, to predict student performance. Such mouse movement trajectories represent the problem-solving path of a student for each question.
Inspired by prior researchers in education and psychology fields~\cite{stahl1994using,row1974wait}, which shows that student's ``think'' time on questions affect their performance,
we propose a set of novel features (e.g., think time, first attempt, first drag-and-drop) based on mouse movement trajectories to predict student performance in interactive online question pools.  Specifically, we define and measure the time between when students see the question and the time that they start solving the questions as the ``think time''. In addition, attributes related to the first attempt and first drag-and-drop are also extracted.
% For example, the features built on students' think time have been proposed.
Further, since there is no predefined question order in interactive online question pools, different students can work on the questions in different orders. We apply a heterogeneous information network (HIN) to calculate the similarity among questions, in an effort to incorporate students' problem-solving history to enhance performance prediction in online question pools.
% incorporating students' historical information to enhance student performance prediction in online question pools.
We evaluated our approach on a real-world dataset that are collected from a K-12 interactive mathematical question pool \textit{Learnlex}. We tested our new feature set on four typical multiple-classification machine learning models -- Logistic Regression (LR), Gradient Boosting Decision Tree (GBDT), Support Vector Machine (SVM) and Random Forest (RF). 

The contributions of this paper can be summarized as follows.

\begin{itemize}
    \item We introduce novel features based on student mouse movement trajectories to predict student performance in interactive online question pools. Features like the ``think time'' can reveal students' thinking details when working on a specific question.
    % Interactive mouse movement trajectories feature regarding the ``think time'' of student can indicate students' thinking details when working on a specific question.
    
    \item We propose a novel approach based on HIN to incorporate students' historical problem-solving information on similar questions into the performance prediction on a new question.
    
    \item We evaluate our approach using real-world dataset and compare with state-of-the-art baseline features on typical multiple-classification machine learning algorithms.
    The 4-class prediction result shows that our approach achieves a much higher performance prediction accuracy than the baseline features in various models, demonstrating the effectiveness of the proposed approach.
    % 76\% which is better than prior studies \cite{o2018student, ren2016predicting}.
\end{itemize}

%% file: src/02_relatedwork.tex
\section{Related Work}
The related work can be categorized into three groups: student performance prediction, problem-solving feature extraction, and question similarity calculation. 
%The related work of this paper can be categorized into xxx groups: xxxx, xxx and xx. 2-pages

\subsection{Student Performance Prediction}
\label{sec_performance_prediction}
There are mainly two ways in performance prediction: the knowledge tracing and the traditional machine learning approach (e.g., Multiple regression). Methods based on or extended from knowledge tracing usually utilize a computational model of the effect of practice on KCs (i.e. knowledge components, which may include skills, concepts or facts) as the way to individually monitor and infer students' learning performance~\cite{piech2015autonomously}. Bayesian Knowledge Tracing (BKT)~\cite{corbett1994knowledge} was the most popular approach to model students' learning process, where each learning concept was represented as a binary variable to indicate whether or not the student has mastered the learning concept or not. However, this method gives little consideration to the individual students' learning ability. Learning Factors Analysis (LFA)~\cite{cen2006learning} extended the basic formulation by adding the factor--learning rate. More factors were taken into consideration by later methods, such as Performance Factors Analysis (PFA)~\cite{pavlik2009performance}, which further incorporated the students' responses to the questions (correct or incorrect). Additive Factors Analysis Model (AFM)~\cite{chounta2019square} added the step duration (time between actions). The performance was improving as more factors are considered. However, the drawback of the methods based on the knowledge tracing is that they need a requirement for accurate concept labeling. Though recent studies showed that there is a possibility to make use of deep learning algorithm (i.e., Recurrent Neural Network) to predict students' performance on consecutive questions, the performance highly relied on the predefined question order (i.e., most of the students followed the same order to solve questions)~\cite{piech2015deep}.

For questions in question pools, they have no predefined order and no accurate concept labels~\cite{xia2019peerlens}, which hinders the way to easily adapt methods based on the knowledge tracing. Many studies have used traditional machine learning methods to predict the drop out rate or the course grade~\cite{manrique2019analysis, chen2016dropoutseer, kabakchieva2013predicting, moreno2018prediction,maldonado2018predicting}. For example, Naive Bayes (NB), RF, GBDT, SVM, k-Nearest Neighbour (KNN) with Dynamic Time Warping (DTW) distance were used to predict college students' dropout~\cite{manrique2019analysis}. Chen~\emph{et al.}~\cite{chen2016dropoutseer} also used Logistic Regression and Nearest-neighbors to predict the dropout in MOOCs. Kabakchieva and Dorina ~\cite{kabakchieva2013predicting} compared different machine learning models (Decision tree classifier, Bayesian classifiers, KNN classifier) for college students for grade prediction. Regardless of the chosen algorithm, much of the performance of a prediction model depended on the proper selection of feature vectors~\cite{manrique2019analysis} and the transformation of feature vectors~\cite{heaton2016empirical}. To achieve a higher accuracy of student performance prediction in the interactive online question pools, we need to extract more meaningful features that are closely related to students problem-solving capabilities.

\subsection{Problem-Solving Feature Extraction}
% \subsubsection{Mouse Movement Feature by other}
Many problem-solving features have been extracted and applied to the student performance prediction on questions. Among them, the submission record and the clickstream are two data types that are studied most. Xia~\emph{et al.}~\cite{xia2019peerlens} used the submission records to model how a student solves a particular problem in online question pools, e.g., solving the problem with one submission or many submissions. Chen and Qu~\cite{qu2015visual, chen2015peakvizor} extracted features of the clcikstream data of watching MOOC videos and analyzed its correlation with final grades. Chounta and Carvalho~\cite{chounta2019square} proposed the use of the response time (i.e., the time between seeing the questions and giving a response to tutor's question or task) to predict students' performance for unseen tasks in intelligent tutoring systems. The result showed that the quadratic response time parameter outperformed the linear response time parameter in the prediction tasks.

However, students' problem-solving behavior is a complex process involving different stages -- problem decomposition, abstraction, and execution~\cite{stahl1994using, yadav2016computational}. It is critical to figure out different stages to better understand and predict their influence on individuals' performance ~\cite{xiong2011analysis} 
%\xm{find another accurate reference for this} 
rather than treating it as a whole solving period. In the interactive question pool we study, more detailed mouse movement sequences (i.e., the trajectories with both position and timestamp information) are collected, which reflect the students' problem-solving ability in extensive details. 
% Inspired by the ``think time'', proposed by Stah~\emph{et al.}~\cite{stahl1994using} as a distinct period of uninterrupted silence by the teacher and all students so that they both can complete appropriate information processing tasks, we attempt to figure out the thinking period from the problem-solving process and extract the thinking time (i.e., the time between viewing the question and having the first mouse click~\xm{please double check}) by using the mouse movement sequences. 
Stahl~\emph{et al.}~\cite{stahl1994using} introduced \textit{``think time''}, which is a period of uninterrupted silence time given by the teacher in class and all students are asked to complete appropriate information processing tasks. Different students may have different abilities in processing the question information before they start to solve it~\cite{thaker2019comprehension}.
%\yong{Huan, what is the purpose and function of ``think time'' in Stah's research? Pls add it here.}
Inspired by these previous work, we propose extracting ``think time'' from students' mouse movement trajectories to delineate the thinking process and problem-solving capabilities of different students.

\subsection{Question Similarity Calculation}
% \subsubsection{other works, typical question similarity method}
Question similarity is one of the key features to infer students' performance in previous researches. For example, When students repeatedly solve questions under the same topic, they may improve the mastery on this learning concept ~\cite{corbett1994knowledge,cen2006learning,pavlik2009performance}. 
% Finding similar questions is a key step in the performance prediction since when there is no interaction logs for the current predicted question, the algorithm should infer students' performance from the similar questions. 
% \yong{Meng, is the above sentence really solid? Pls check if the claim is true.}
A wide range of work has been done to calculate question similarity when there is no accurate expert annotation of the problems. One branch of work calculated the semantic similarity of questions based on Natural Language Processing (NLP). For example, Song~\emph{et al.}~\cite{song2007question} first identified keywords in questions, and then calculated semantic similarity between questions based on the keywords. They also extended the semantic similarity to statistic similarity, which was calculated using the cosine similarity between two question vectors. Each question vector was a string of binary bits with each bit indicating the existence of a certain word.
% in the words set of two question. 
Their results showed that the combination of semantic similarity and statistic similarity could achieve a better performance than each individual algorithm. Similarly, textual similarities~\cite{charlet2017simbow} and question type similarity~\cite{achananuparp2008utilizing} were also proposed to further improve the performance prediction.

% Statistic 50.0\%
% Semantic 57.1\%
% Combined 64.3\%

% semantic:
% syntactic:
% Semantic and syntactic information is measured by taking into account word similarity, word ordering, and parts of speech information. 

% question-type similarity: We employ the information about the types of questions, provided by a trained text classifier, to further differentiate similar/dissimilar questions. 
% ~\cite{achananuparp2008utilizing}

% This work~\cite{charlet2017simbow} calculated textual similarities, which relied on the introduction of a relation matrix in the classical cosine similarity between bag-of-words, so as to get a softcosine that takes into account relations between words.
% Existing methods NLP
% \subsubsection{Heterogeneous Information Network and Similarity Calculation}
%The title of this subsection can also be changed to some other alternatives, sth. like, ``Question Similarity Evaluation in Online Question Pool''.

% To calculate the similarity of questions with text information, an algorithm based on semantics of sentences and statistic of words \cite{Achananuparp_utilizingsentence}. Besides, some researchers proposed methods to calculate similarity of math expressions \cite{zhang2014math, Shahab2013structural}, which could also be used to search similar questions. 

However, all these methods require accurate knowledge tags and abundant text information about the questions. The questions in the interactive math question pools usually use simple description to describe the problem background without showing the knowledge tested explicitly. Another branch of work tried to calculate the question similarity according to the interaction data between students and problems. For example, Xia~\emph{et al.}~\cite{xia2019peerlens} calculated the question similarity based on the submission types (e.g., successful submission after one submission or many submissions). 
% (how students finish the questions, with one attempt or many attempts). 
% They did not combine more detailed features extracted from the mouse movement sequences. 
However, more detailed student interaction information (e.g., mouse movement trajectories), which intrinsically reflects students' problem-solving habits and capabilities, is not considered.
HIN is defined as the information network with more than one type of objects or relationship between objects, which is used to describe the complex structure of information in the real world. Bibliographic information network
% built on the dataset of DBLP
and Twitter information network are examples of HIN \cite{sun2013mining}. In this paper, we use the HIN to incorporate different information (e.g., students' historical scores, mouse movement trajectories) %(e.g., \xm{what kinds of information are used in HIN}) 
into calculating the question similarity.
% or math expressions is required to calculate similarity so they cannot be applied on questions with simple text description and various graphic information.

% According to Shi et al.\cite{shi2017survey}, HIN can be further classified as 4 typical types, multi-relational network with single-typed object, bipartite network, star-schema network and multiple-hub network. Previous works on bipartite network have modeled some relationship including term-document \cite{rossi2012bipartite}, terrorist-event \cite{Gollini2019AMM} and user-item \cite{Jamali2013recommend}. 

% However, there hasn't been any work which used bipartite HIN to describe the relationship between students and questions.

%\yong{Overall, only four or five papers are cited in Sections 2.2 and 2.3. It would be better to include more related papers. }

%% file: src/03_background.tex
\section{Context}
% In this section, we introduce the basic information of our dataset and the prediction framework.
% \subsection{Data description}
Our study is built on the dataset collected from an interactive online math question pool. This section introduces the interactive math questions, the collected data and the overall student performance prediction framework of our study.

\subsection{Interactive Math Question}
\begin{figure}[h]
    \centering
    \includegraphics[width=0.7\linewidth]{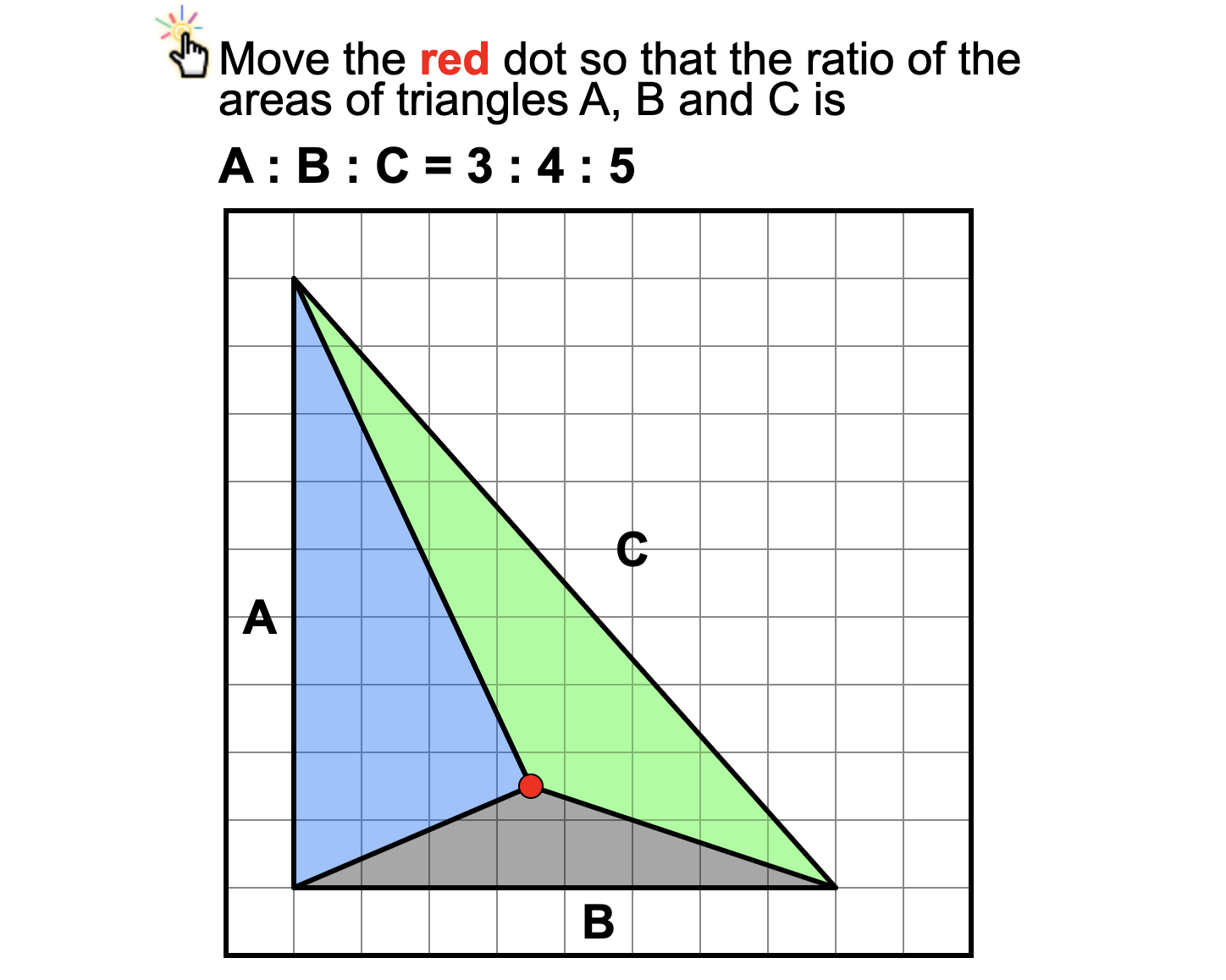}
    \vspace{-1em}
    \caption{An example of the interactive math question.}
    % \caption{An Example of interactive question(Area dimension).}
    \label{example}
\end{figure}

The platform we cooperate with is an interactive question pool used by more than 40,000 students from 30+ primary and junior high schools in Hong Kong. There are 1,720 interactive math questions on the platform and each question has labels from \textit{math dimension}, \textit{difficulty}, and \textit{grade}. The \textit{math dimension} indicates the general knowledge domain of the question. \textit{Difficulty} is a five-star rating ranging from 1-5 (easy to hard), which is predefined by several education experts and the question maker. \textit{Grade} represents the year of study this question is designed for. Figure~\ref{example} shows an example of the interactive math question in the question pool, students need to use their mouse to drag the red dot to fulfill the requirement and get a score ranging from 0 to 100. Since the scores are discrete and possible scores of each question are not the same, we manage to map the original score ranging from 0 to 100 to 4 score classes (0-3).

\subsection{Data Collection}
The data we collected are historical score records and mouse movement trajectories. In total, we collected 858,115 historical score records from September 13, 2017 to June 27, 2019 in total. We developed a tool to collect students' mouse movement trajectories, which included the mouse events (mouse move, mouse down, mouse drag and mouse up), the timestamps, and the positions of the mouse during the whole problem-solving process. We selected two question sets that contains rich mouse interactions from two math dimensions (\textit{Area} and \textit{Deductive geometry}).
% \yong{It is better to provide some basic explanations about why we choose this two set of questions. How about the scoring difference (i.e., 4-class vs. xx-class ?)?}
Table \ref{tabledataset} shows the statistical information of these two question sets. There are 61 and 64 questions in these two categories respectively. 724 students produced 1764 mouse movement trajectories in the Area and 562 students made 2610 mouse movement trajectories in Deductive geometry. Note that the system allows students to submit up to two times for each question, our research goal in this paper is to predict students' score on the first submission of the question.

% We will conduct experiments on the two question set respectively to make sure the generalization of our features and methods.
%and questions have same math dimension in each question set. 
% \usepackage{booktabs}
\begin{table}[]
\begin{tabular}{@{}cccc@{}}
\toprule
Dataset & \#Questions & \#Trajectories & \#Students \\ \midrule
ADD & 61 & 1764 & 724 \\
DGDD & 64 & 2610 & 562 \\ \bottomrule
\end{tabular}
\caption{The statistics of two datasets: ADD (area dimension dataset) and DGDD (deductive-geometry dimension dataset). Both consist of students' problem-solving records from April 12, 2019 to June 27, 2019.}
\label{tabledataset}
\vspace{-3em}
\end{table}

\subsection{Overall Prediction Framework}
\begin{figure}[h]
    \centering
    \includegraphics[width=8cm]{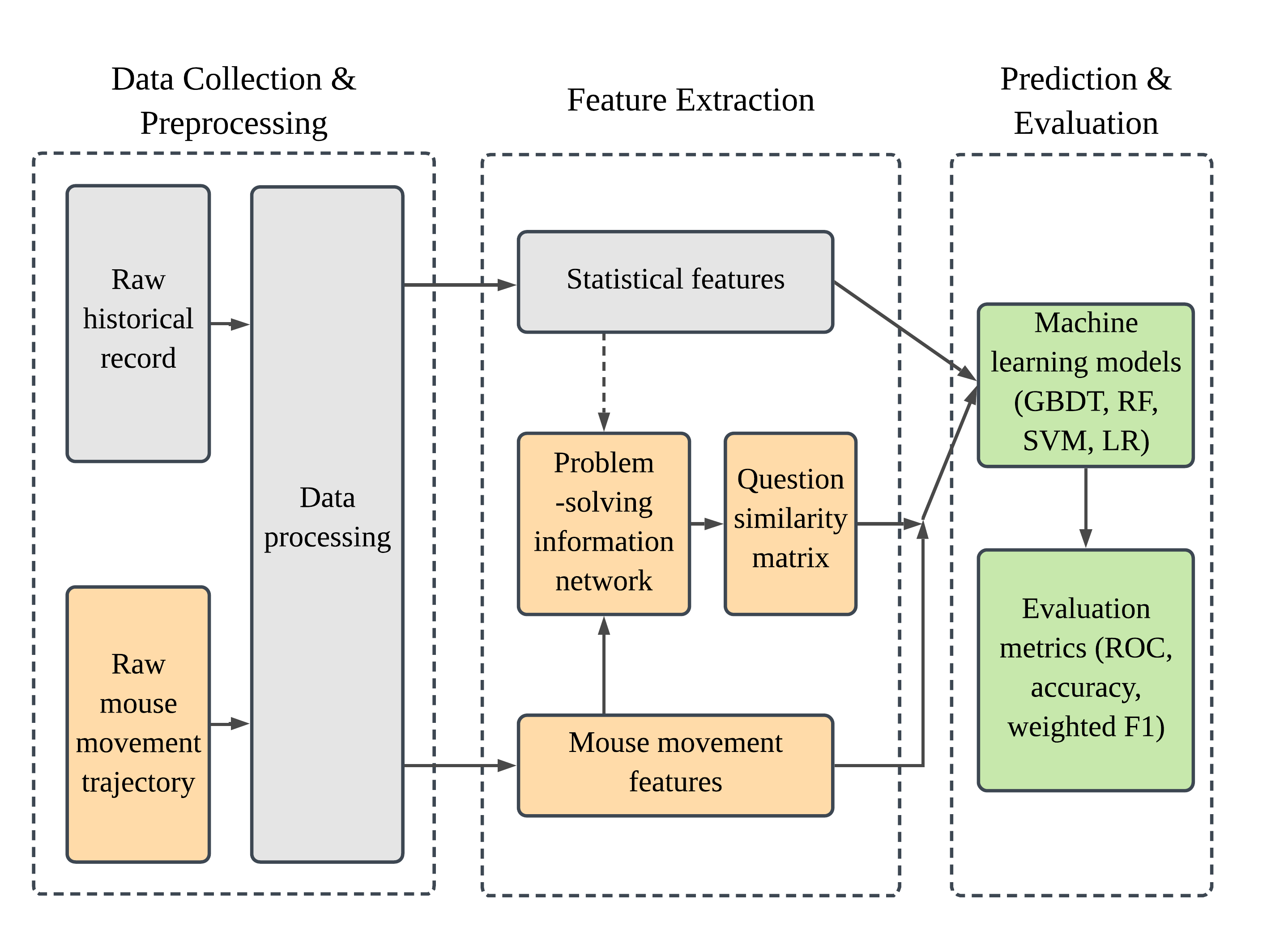}
    \vspace{-1.5em}
    \caption{The prediction framework contains three modules: data collection \& preprocessing module, feature extraction module, and prediction \& evaluation module. The blocks highlighted in yellow and green correspond to the major contributions of this paper.}
    \label{framework}
\end{figure}

% In this section, we will discuss each part of figure 2
% \begin{itemize}
%   \item Raw historical data contains question sets' statistics and students' historical records from Apr. 12th, 2019. We extract the features and The corresponding processed data is statistical features.
%   \item Mouse movement sequences 
% \end{itemize}
In order to conduct a state-of-the-art machine learning prediction experiment, we survey existing studies for
the features used in student performance prediction~\cite{o2018student, manrique2019analysis},
mouse trajectory feature definition~\cite{yamauchi2013mouse,seelye2015computer},
and possible methods in question similarity calculation ~\cite{song2007question,charlet2017simbow,achananuparp2008utilizing,sun2013mining,sun2011pathsim}.
% and classification models \cite{heaton2016empirical,}. ~\cite{,gardner2019evaluating}
We summarize this knowledge with our dataset and task, then we propose our prediction framework in Figure~\ref{framework}. It mainly contains three modules: data collection and preprocessing, feature extraction, and prediction and evaluation. 

% Specifically, in the data collection and preprocessing module, we collect the students' past score records and process them into two parts -- historical and recent records. 
In the feature extraction module, as suggested by previous studies~\cite{galyardt2014recent,goldin2015convergent}, students' recent performance may have a great impact on prediction. We then extract the historical performance statistical features and recent performance statistical features from the records. In addition, we summarize each question's basic information (e.g., grade, difficulty) as well as the number of total submissions and second submissions.
%\xm{decribe what you did in this step in detail}.
% As for mouse movement trajectories, we collect mouse events, the timestamp, and the mouse position.  %%%%ALREADY said.
% . However, all the questions are based on drags and mousedrag is raw mouse move event between mousedown and mouseup (Figure\ref{fig4}), 
% We then process the raw trajectories into mouse move and drag-and-drop. 
% In the feature extraction module, 
Further more, mouse movement trajectories are processed to representation features including think-time, first drag-and-drop, first attempt, and other problem-solving related features. 
In addition, we combine these features and statistical features (e.g., score) and build the problem-solving information network, a HIN, on them to calculate the question similarity matrix.

In the prediction and evaluation module, we test these features on four typical machine learning models to compare their performances with and without the mouse movement features. Similar to previous research ~\cite{gardner2019evaluating, manrique2019analysis}, we compare prediction accuracy, weighted F1 and ROC curves and feature importance score in GBDT to evaluate and discuss our method.
%Deductive geometry is about angles and lines.
%Area is about parallelograms, triangles, trapeziums and polygons. 
% There are 61 questions in area dimension question sets and 64 questions in deductive-geometry dimension question sets
%Here is an example of a student's mouse movement in figure 3, each dot represents a mouse event triggered by mouse and 
% He moves mouse from A to B and drags from B to C and so on. 
% Students solve the math questions by their ideas. 
%We defined and extract several features from mouse movement sequence.
%Maximization of the likelihood is relatively simple and is commonly performed using expectation maximization [9], bruteforce grid search [1] or gradient descent [42].
%Including question types, difficulty and students' historical score, we also collect student's question-solving data -- mouse movement, drag and drop sequence on each question.
% \subsection{Question Labels}
% \subsection{Historical Data}
%Historical data is student's historical performance of the questions they solved including score, time, try times on questions.
% \subsection{Mouse Event}
%Mouse events including movement, mouse down and mouse up are collected in web page. Mouse drag not a prototype mouse event in DOM[], it is a series of constant mouse move event starting with mouse down and ending with mouse up. 

%% file: src/04_method.tex
\section{Feature Extraction}
In this section, we introduce the feature extraction module in detail. Specifically, we first explain how to detect change points in the mouse movement trajectories, based on which we illustrate how we extract mouse movement features
(i.e., think time, first drag-and-drop, and first attempt).
% \yong{Huan, we need to give an explicit definition or introduction for the three concept in some parts of Section 4.1 or 4.2. Pls add it in an appropriate place.}
% Drag-and-drop can be considered as a series of mouse drag events that start with the mouse down and end with the mouse up. First attempt is the first action episode after the think time, which might be the real beginning of problem-solving interaction.
% \yong{what is ``first action''?}
% Statistic measurements(e.g., mean, median) are applied to extract drag features in each mouse movement trajectory.
We then introduce other statistical features like students' historical performance features. Lastly, we describe how to use a HIN to incorporate both statistical features and mouse movement features to calculate the similarities between questions. Assuming that students often have similar performance on similar questions, we integrate features of similar questions to further enhance the performance prediction on a new question.
% In our program framework, we need to have the student's mouse movement data on the candidate question before he or she finishing it, so finding a similar question to the candidate one is a key process in our task. In general situation, we use similarity method(e.g., cosine similarity) and question's basic information such as difficulty, score, grade or other traditional features to calculate similarities. 
\begin{figure}[h]
    \centering
    \includegraphics[width=8cm]{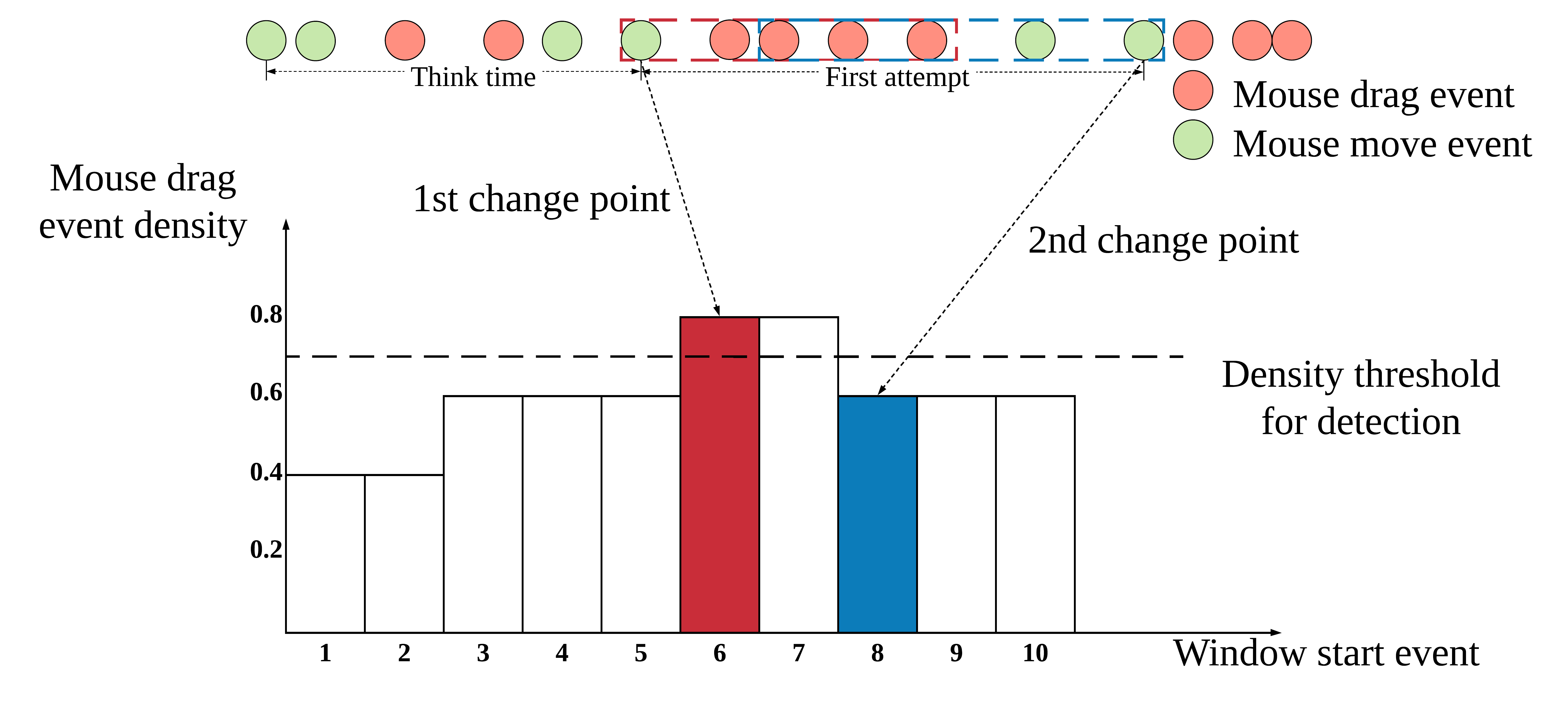}
    \vspace{-2em}
    \caption{A sample mouse movement trajectory and the scheme diagram of change point detection algorithm. The 1st change point is both the think time end point and the first attempt start point. The 2nd change point is the first attempt end point.}
    \label{change point}
    \vspace{-1em}
\end{figure}
% we utilize Heterogeneous Information Network (HIN) and mouse movement features to calculate a question similarity matrix. 
% %\xm{ADD reason}
\subsection{Change Point Detection}
\label{sec_change_point}
To infer students' problem-solving capabilities from the mouse movement trajectories, we need to identify different problem-solving stages~\cite{yadav2016computational,xiong2011analysis}. % For example, the think time stage is a period for reading and understanding the question before solving it with actual actions such as the mouse drag. 
% \yong{We need some sentences to explain the relationship between change points and the student performance. References are needed to support the claims.}
The change points are the time points where there is an abrupt change in the mouse movement trajectory to distinguish different problem-solving stages in our context.
As shown in Figure~\ref{change point}, we split the mouse movement trajectory into three subparts using two change points: the think time stage, the first attempt stage (i.e., the first action episode after the thinking period) and the following actions stage. Think time is a stage that starts from when a student opens the question and ends at the students use the mouse to solve the question with actual interactions~\cite{stahl1994using}. The first attempt stage is a series of mouse drag events trying to solve the problem after the think stage and ends when the frequency of mouse drag events becomes low. The first change point can differentiate the think time stage from the first attempt stage and the second change point is the boundary of the first attempt stage and the following actions stage. 
% Think time\cite{} can reflect a student' pause period from the time they open the question to the time they start to solve the question with the first actual action. 
A straightforward way of change point detection is to use the first movement of the mouse as a signal that the student starts to solve the question. However, in the real world, it is not always true that the first mouse movement represents the starting point of solving a problem. After analyzing the mouse movement trajectories, we find that there may exist a short drag-and-drop or click by mistake, which makes it seem like the student starts to solve questions. To detect the most probable start and end time points of each stage, we propose the change point detection algorithm.
\vspace{-2em}
\begin{figure}[h]
    \vspace{-1em}
    \centering
    \includegraphics[width=9cm]{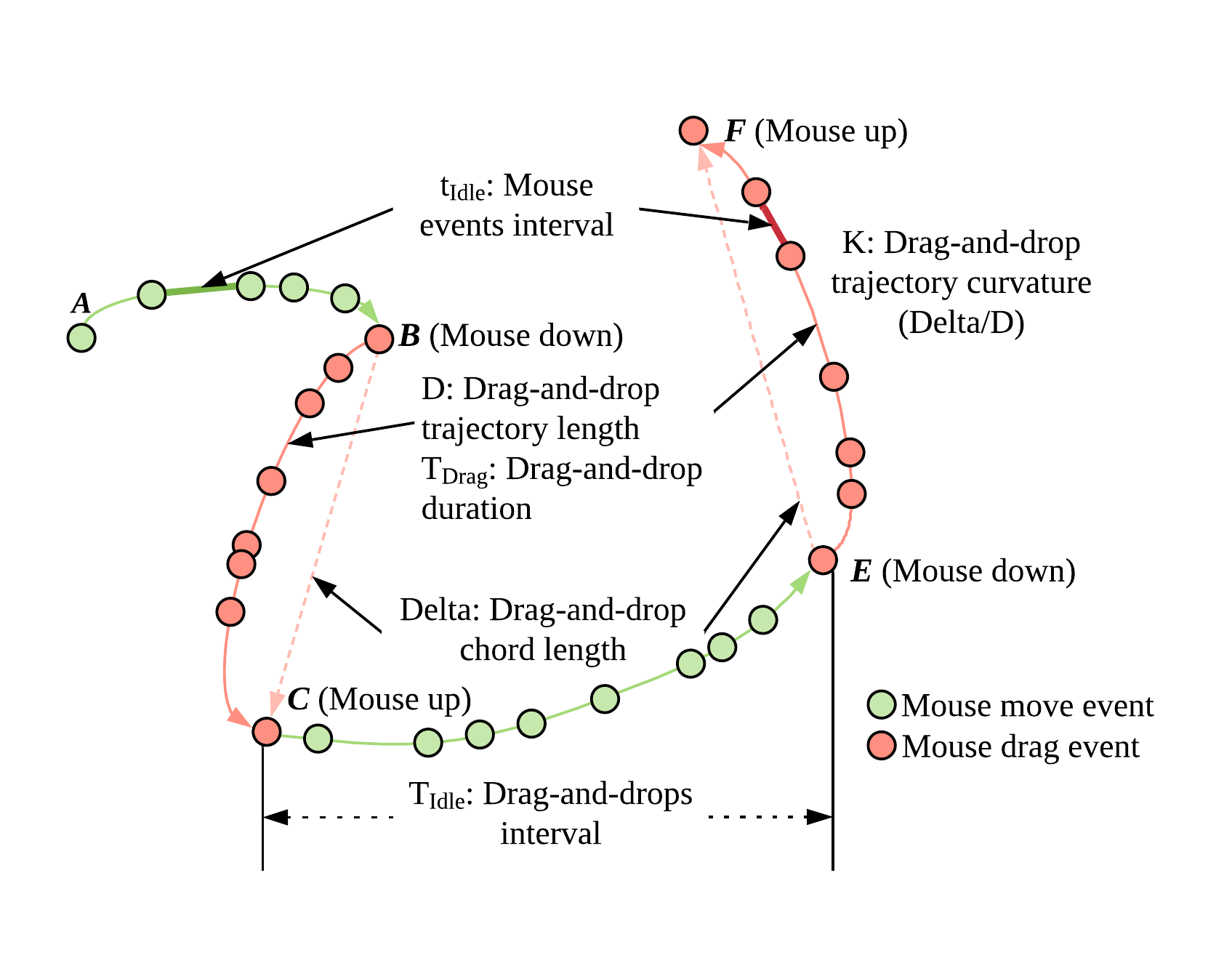}
    \vspace{-4em}
    \caption{An example of mouse movement trajectories in problem-solving process.}
    \label{trajectory}
    \vspace{-1em}
\end{figure}
% \subsubsection{Sliding window method}
% As shown in Figure~\ref{change point}, we split the mouse movement trajectory into three subparts: the think time stage, the first attempt stage (i.e., the first action episode after the thinking period) and the following actions stage with two change points. The first change point can differentiate the think time stage from the first attempt stage  and the second change point is the boundary of the first attempt stage and the following actions stage. The time between the moment when a student opens the question and the first change point is the think time~\cite{stahl1994using} and the action between the first change time and the second change time is the first attempt.

We use the sliding window to detect the change points, which is a common method for abrupt change detection in the time series data~\cite{basseville1993detection}. It has two key parameters: window size and detection threshold.
% They varies in different situations. 
The detection threshold can be observed through the test data or sometimes random selection~\cite{basseville1993detection}. 
% \subsubsection{Density threshold} 
For a mouse movement trajectory, we define the total mouse events as $C_t$, the total mouse drag events as $C_d$. The event density in trajectory is defined as:
\begin{equation}
    \rho = C_d/C_t
\end{equation}
% \subsubsection{Change point detection}
We count the mouse drag events and calculate the mouse drag event density in sliding window and compare the value of density with the threshold. The scheme of change detection is as follows:
\begin{itemize}
% \item When the first time $\rho$ is higher than threshold, we define the first point of the window as a change point of think time, which is also the start point of the first attempt.
% \item When think time change point is detected, continue sliding window until $\rho$ is lower than threshold. We define the last point of the window as the change point of first attempt ends.
\item Move the window from the beginning of the mouse events sequence, when the first time $\rho$ is higher than the threshold, the first point of the sliding window is defined as the first change point.
\item Continue sliding window until $\rho$ is lower than the threshold, the last point of the sliding window is defined as the second change point.
% \yong{I cannot understand it.}
\end{itemize}
%If not, move the window one event forward and continue until it passes the threshold. Then, we set the first point as the think time end point, 
\begin{table}[]
\small
\begin{tabular}{@{}lll@{}}
\toprule
Type & Feature & Description \\ \midrule
\multirow{4}{*}{Think time} & Time length & \begin{tabular}[c]{@{}l@{}}Time between opening the web \\ page and the first change point.\end{tabular} \\ \cmidrule(l){2-3} 
 & Time percent & \begin{tabular}[c]{@{}l@{}}Percentage of think  time in \\ time length of the whole trajectory.\end{tabular} \\ \cmidrule(l){2-3} 
 & Event length & \begin{tabular}[c]{@{}l@{}}Mouse event number \\ in think time.\end{tabular} \\ \cmidrule(l){2-3} 
 & Event percent & \begin{tabular}[c]{@{}l@{}}Percentage of mouse event\\ number in the whole trajectory.\end{tabular} \\ \midrule
    First attempt & \begin{tabular}[c]{@{}l@{}}Event end \\ index\end{tabular} & \begin{tabular}[c]{@{}l@{}}Mouse event number during\\ the first attempt.\end{tabular} \\\midrule
\multirow{9}{*}{\begin{tabular}[c]{@{}l@{}}First drag-\\ and-drop\end{tabular}} & Time length & \begin{tabular}[c]{@{}l@{}}Time between opening the web \\ page to the first mouse drag.\end{tabular} \\ \cmidrule(l){2-3} 
 & Time percent & \begin{tabular}[c]{@{}l@{}}Percentage of first drag-and-drop\\ in time length of the whole trajectory.\end{tabular} \\ \cmidrule(l){2-3} 
 & \begin{tabular}[c]{@{}l@{}}Event start\\ index\end{tabular} & \begin{tabular}[c]{@{}l@{}}Mouse  event  number  before\\ first drag-and-drop starts.\end{tabular} \\ \cmidrule(l){2-3} 
 & Event percent & \begin{tabular}[c]{@{}l@{}}Percentage of event number \\ before first drag-and-drop starts\\ in the whole trajectory..\end{tabular} \\ \cmidrule(l){2-3} 
 & \begin{tabular}[c]{@{}l@{}}Event end\\ index\end{tabular} & \begin{tabular}[c]{@{}l@{}}Mouse event number when first\\ drag-and-drop ends.\end{tabular} \\ \cmidrule(l){2-3} 
 & K & First drag-and-drop trajectory curvature. \\ \cmidrule(l){2-3} 
 & D & First drag-and-drop trajectory length. \\ \cmidrule(l){2-3} 
 & Delta & First drag-and-drop chord length. \\ \bottomrule
\end{tabular}
\caption{TFF feature table. TFF: think time, first attempt, and first drag-and-drop.}
\label{table:3}
\vspace{-3em}
\end{table}
\subsection{Mouse Movement Features}
Student interaction data, especially mouse movement data, contains massive information~\cite{hagler2014assessing,seelye2015computer}.
We extract two sets of features: TFF (think time, first attempt, and first drag-and-drop) and MDSM (mouse drag statistical measurements) based on previous studies~\cite{stahl1994using,yamauchi2013mouse,seelye2015computer}.
We use TFF and MDSM to model students' initial and overall behaviors in a problem-solving process, respectively.

As for the TFF, the think time and the first attempt have been introduced in Section 4.1. Drag-and-drops is a series of consecutive mouse drag events that start with the mouse down and end with the mouse up and thus the first drag-and-drop is the first drag event with a mouse down and a mouse up. Table~\ref{table:3} shows the detailed attributes of the features in TFF.

As for MDSM, we define the following features to represent mouse drag statistical measurements as Figure 4 shows.
\begin{itemize}
    \item $D$: Drag-and-drop trajectory length.
    \item $Delta$: Drag-and-drop chord length.
    \item $K$: Drag-and-drop trajectory curvature (Delta/D).
    \item $T_{Drag}$: Drag-and-drop duration.
    \item $T_{Idle}$: Drag-and-drops interval.
    \item $t_{Idle}$: Mouse events interval.
\end{itemize}
 
However, there may be more than one drag-and-drop in a trajectory. To extract useful information in drag-and-drops, we use three statistical methods median, mean and interquartile range (IQR), to measure it (Table ~\ref{tableMmeasure}). Besides the features above, we also build other features to measure the whole trajectory, including the number of drag-and-drops, total time, total mouse events and mean, median, IQR of mouse events in every second.

% To extract useful information in mouse drags in movement trajectories 
% As shown in Figure~\ref{trajectory}, $t_{Idle}$ and $T_{Idle}$ are used to represent the interval between mouse events (i.e., mouse move, mouse up, and mouse down) and the interval between drag-and-drops, respectively. 
% We  in drag-and-drops (Table~\ref{tableMmeasure}).

% to dig out useful information in the whole mouse movement trajectory. We list The detailed explanation of these features is listed below.

% \usepackage{booktabs}
\begin{table}[]
\small
\begin{tabular}{@{}clcccccc@{}}
\toprule
 & \multicolumn{1}{c}{$K$} & $D$ & $T_{drag}$ & $Delta$ & $t_{Idle}$ & $T_{Idle}$ &\\ \midrule
Median & \multicolumn{7}{l}{The feature's median value in its value list of the whole trajectory.} \\
IQR & \multicolumn{7}{l}{The feature's IQR in its value list of the whole trajectory.} \\
\multicolumn{1}{l}{Mean} & \multicolumn{7}{l}{The feature's mean value in its value list of the whole trajectory.} \\ \bottomrule
\end{tabular}
\caption{Three statistical measures for six features. IQR: Interquartile range. $D$: Drag-and-drop trajectory length,  $Delta$: Drag-and-drop chord length, $K$: Drag-and-drop trajectory curvature, $T_{drag}$: Drag-and-drop duration,  $T_{Idle}$: Drag-and-drops interval and $t_{Idle}$: Mouse events interval.}
\label{tableMmeasure}
\vspace{-3em}
\end{table}

\subsection{Other Statistical Features}
% Despite mouse movement trajectories are important,
% \yong{We should also add sentences to explain these features are widely used. References are also needed.}
Information such as score records and score distribution on questions can also reflect a student's ability and the question difficulty for all the students. Features extracted from such records are widely applied in prediction of students' performance, for example, Yu \textit{et al.}~\cite{yu2010feature} used students' recently solved questions to construct temporal features and Manrique \textit{et al.} introduced average grades of students to predict their dropout rates~\cite{manrique2019analysis}.
To make full use of the information, we take the cross feature approach. A cross feature is a synthetic feature formed by multiplying (crossing) two or more features. Crossing combinations of features can provide predictive abilities beyond what those features can provide individually~\cite{crossfeature}. Thus, we apply the cross feature method to questions' and students' basic statistics. As Table~\ref{table_other_fea} shows, we have three parts of statistical features: question statistics, student statistics, and the recent statistics of a student. 

We use the expression A $\times$ B to represent the cross feature of A and B, expression \#C in $[$ E $]$ to represent the numbers of C for each category or dimension in E and expression \%C in $[$ E $]$ to represent the proportion of C for each category or dimension in E. For example, in students statistics, \%Submission in $[$math dimension $\times$ grade $\times$ difficulty$]$ represents the proportion of a student's submission number in each math dimension with a specific grade and difficulty level.
\begin{table}[]
\small
\begin{tabular}{l|l}
\hline                                                                            Feature & Description                                               \\ \hline
\multicolumn{2}{c}{\textbf{Question statistics}}                            \\ \hline \hline
% \multirow{6}{*}{\begin{tabular}[c]{@{}l@{}}Question\\ statistics\end{tabular}}          & 
Math dimension                                                                                                                                        & Question's domain knowledge (e.g, area).                                                                                                                                         \\ \hline
                                                                                        Grade                                                                                & \begin{tabular}[c]{@{}l@{}} Student's grade that the question suggests.\end{tabular}  \\ \hline
                                       Difficulty                   & \begin{tabular}[c]{@{}l@{}}Question's difficulty given by experts.\\ \end{tabular}   \\ \hline
                                 \#Total submissions                         & \begin{tabular}[c]{@{}l@{}}Total number of submissions in question.\end{tabular}                                                                                     \\ \hline
                                                                                        \#2nd submissions                                                                                                                                & \begin{tabular}[c]{@{}l@{}}Total number of second submissions in question.\end{tabular}                                                                                 \\ \hline  \begin{tabular}[c]{@{}l@{}}
                                                                                       \%Submissions in \\ $[$score class$]$
                       \end{tabular}    
                                    &                 \begin{tabular}[c]{@{}l@{}}Proportion of submissions in each score class.\end{tabular}                                                                                                     \\ \hline \hline
% \multirow{3}{*}{\begin{tabular}[c]{@{}l@{}}Student\\ statistics\end{tabular}}           & 
\multicolumn{2}{c}{\textbf{Student statistics}} \\ \hline \hline
\#Total submissions                                                                                                                                & \begin{tabular}[c]{@{}l@{}}Student's total submissions in history.\end{tabular}                                                                              \\ \hline
                                              \#2nd submissions                                                        & \begin{tabular}[c]{@{}l@{}}Student 's total second submissions in history.\end{tabular}                                                                     \\ \hline \begin{tabular}[c]{@{}l@{}} \%Submissions in \\$[$math dimension $\times$ \\ grade $\times$  difficulty$]$\end{tabular}                          & \begin{tabular}[c]{@{}l@{}}Student's proportion of submissions in each \\specific math dimension, grade and \\ difficulty.\end{tabular}                   \\ \hline \begin{tabular}[c]{@{}l@{}}  1stAvgScore in \\ $[$math dimension $\times$ \\ grade $\times$ difficulty$]$
                       \end{tabular}                & \begin{tabular}[c]{@{}l@{}}Student's first submission average score in \\ each specific math dimension with assigned \\specific grade and difficulty.\end{tabular}       \\ \hline \hline

% \multirow{6}{*}{\begin{tabular}[c]{@{}l@{}}Student \\ recent\\ statistics\end{tabular}} & 
\multicolumn{2}{c}{\textbf{Student recent statistics}} \\ \hline \hline
\begin{tabular}[c]{@{}l@{}}\#Submissions in \\ $[$math dimension$]$ \end{tabular}                                          &
                                    \begin{tabular}[c]{@{}l@{}}Number of submissions in each math dimension \\in past N days.\end{tabular} \\ \hline 
                                     \begin{tabular}[c]{@{}l@{}}\#Submissions in \\ $[$grade $\times$ difficulty$]$\end{tabular}                                    & \begin{tabular}[c]{@{}l@{}}Number of submissions in each grade and \\difficulty in past N days. \end{tabular}        \\ \hline
                                                                     \begin{tabular}[c]{@{}l@{}}Average score in \\ $[$math dimension$]$\\ \end{tabular}                        & \begin{tabular}[c]{@{}l@{}}Average score in each math dimension in past\\ N days.\end{tabular}                       \\ \hline 
                                                                     \begin{tabular}[c]{@{}l@{}}Average score in \\ $[$grade $\times$ difficulty$]$\end{tabular}                  & \begin{tabular}[c]{@{}l@{}}Average score in each grade and difficulty in\\ past N days.\end{tabular}                  \\ \hline 
                       \begin{tabular}[c]{@{}l@{}}Score std in \\ $[$math dimension$]$ \end{tabular}          & \begin{tabular}[c]{@{}l@{}}Score standard deviation of each math \\ dimension in past N days.\end{tabular}       \\ \hline
                                                                    \begin{tabular}[c]{@{}l@{}}Score std in\\ $[$grade $\times$ difficulty$]$\end{tabular} & \begin{tabular}[c]{@{}l@{}}Score standard deviation of each grade and \\ difficulty in past N days.\end{tabular} \\ \hline
\end{tabular}
\caption{Other statistical features: questions statistics, student statistics and student recent statistics. Symbol: \#: Number of records, \%: Proportion of records. Expression A $\times$ B: cross features of A and B. Expression C in $[$ E $]$: calculate C for each category or dimension in E.}
\label{table_other_fea}
\vspace{-3em}
\end{table}

% All extracted other statistical features are listed in Table \ref{table_other_fea}
\subsection{Problem-solving Information Network and Similarity Calculation} \label{QIN}
To predict the performance of a student $s_i$ on a question $q_x$ using mouse movement features, an intrinsic requirement is that we should have the mouse movement trajectories. 
However, they are not available before 
% we have no real mouse movement trajectory before
a student actually finishes the question.
Prior studies~\cite{yera2018recommender, sanchez2017case} have shown that a student's performances on similar questions are often similar.
Therefore, we propose finding a question $q_y$ similar to the question $q_x$ and
% Thus, we need to find a similar question $q_y$ and 
using its mouse movement features extracted from the trajectories as part of the feature vectors to predict a student's performance on $q_x$.

The similarity between questions can be evaluated from different perspectives (e.g., difficulty level, question content, student mouse movement interactions, etc.). For example, 
% A straightforward wat to check whether the two questions test the same knowledge by using the text descriptions. However, interactive online question pools often avoid revealing such kind of information for examination purpose~\cite{xia2019peerlens}.
for two questions that examine the same knowledge, their difficulty levels and the required problem-solving skills can be significantly different for different students.  To more accurately delineate the similarity between any two questions, we use students' interactions (e.g., mouse movement trajectories) as the bridge from one question to another question. Specifically, we build a network consisting of both interactions between students and questions and the intrinsic attributes of questions to calculate similarity for each pair of questions. Such a network is called \textit{problem-solving information network} in this paper.
% For the similarity evaluation, we consider both a student's score on a specific question and his/her detailed mouse movement trajectories to comprehensively evaluate the question similarity details.
% we construct a \textit{question-student-question} meta-path to 

\iffalse
% Finding the similar question $q_y$ is challenging in our dataset since there is neither enough text information to describe the question nor well pre-defined knowledge components of questions to use the previous methods in Section 2.3. 
The only valuable information can be used to calculate question similarity other than the rough math dimension and score distribution about questions is the mouse movement trajectory data of different students under our scenario.
Moreover, these trajectories involve both students and questions so they cannot be directly used to compute the similarity matrix of questions.
Thus, we use students as the bridge from one question to another question and build a network between students and questions to calculate similarity for each pair of questions.
% In our method, we design and implement the question-solving network for calculation of similarity between every pair of questions. 
Base on the similarity score, we then use the mouse movement features of the most similar question as part of the features of the target question for prediction.
\fi

\subsubsection{Problem-solving Information Network Structure}
\begin{figure}[h]
    \centering
    \includegraphics[width=8cm]{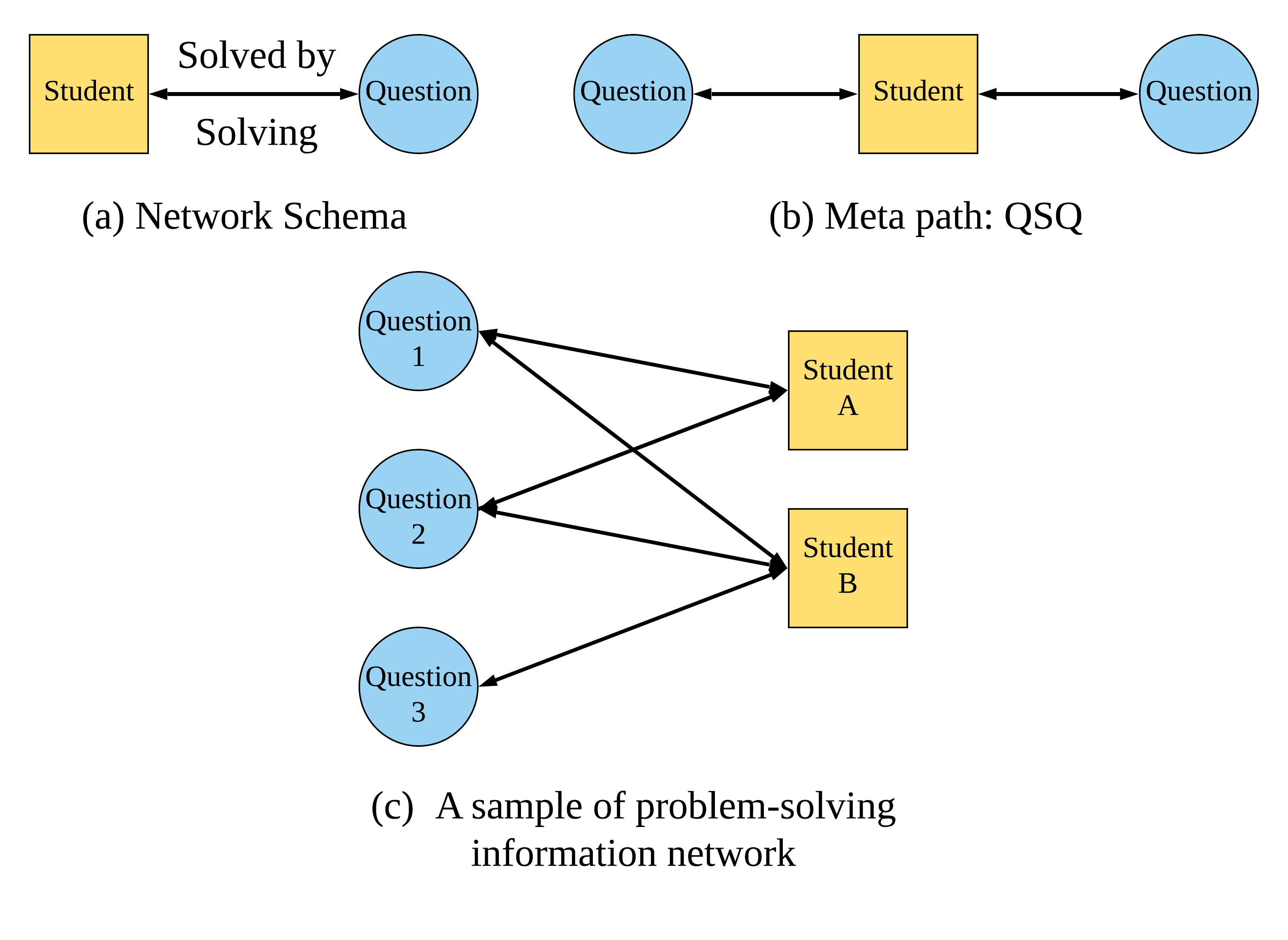}
    \vspace{-2em}
    \caption{The schematic diagrams of problem-solving information network. (a) Network schema, (b) Meta path: $QSQ$, and (c) A sample of problem-solving information network.}
    \label{fig1}
\end{figure}

The problem-solving information network, a typical bipartite HIN, is established between two kinds of objects, students (S) and questions (Q). For each question $q\in Q$, it has links to several students and the link between a question $q$ and a student $s$ is defined as solving ($s$ solves $q$) or solved by ($q$ is solved by $s$). The network schema of the problem-solving network is shown in Figure~\ref{fig1} (a) and the symmetric meta-path in our network is defined as Question-Student-Question ($QSQ$), which denotes the relationship between two questions that have been solved by the same student.

% \vspace{-1em}

\subsubsection{Similarity Calculation}
% PathSim \cite{sun2011pathsim} is an algorithm proposed to retrieve top K similar objects of the same type in a heterogeneous information network based on symmetry meta-paths. We only adapt its measure to calculate similarity score on one meta-path as Equation \ref{eq1} \cite{sun2011pathsim} shows.

% \begin{equation}\label{eq1}
% s(q_x, q_y)=\frac{2 \times\left|\left\{p_{q_x \leadsto q_y} : p_{q_x \sim q_y} \in \mathcal{P}\right\}\right|}{\left|\left\{p_{q_x \leadsto x} : p_{q_x \leadsto q_x} \in \mathcal{P}\right\}\right|+\left|\left\{p_{q_y \sim q_y} : p_{q_y \leadsto q_y} \in\mathcal{P}\right\}\right|}
% \end{equation}

% In Equation \ref{eq1}, $q_x$ and $q_y$ represent two different questions and $\mathcal{P}$ is a meta-path from $q_x$ to $q_y$. In order to combine the similarity scores of several different meta-paths which pass through a group of students, we use the multiple meta-path combination method \cite{sun2011pathsim} as Equation \ref{eq2}.

% \begin{equation}\label{eq2}
% s_{\operatorname{comb}}\left(q_x, q_y\right)=\sum_{l=1}^{r} w_{l} s_{l}\left(q_x, q_y\right)
% \end{equation}

% In Equation \ref{eq2}, the weight $w_l$ of each path $l$ in our model is defined as the cosine similarity score of two mouse movement feature of a student on $q_x$ and $q_y$ divided by the total number of meta-paths between $q_x$ and $q_y$. Thus, our similarity scores' range is limited to $(0, 1)$.

Sun \textit{et al.}~\cite{sun2011pathsim} proposed a meta-path based similarity framework and the framework on meta path $\mathcal{P}$ is defined as:
\begin{equation} \label{eq_sim1}
    s(a, b)=\sum_{p \in \mathcal{P}} f(p)
\end{equation}
% In Equation~\ref{eq_sim1},
where $f(p)$ denotes a measure defined on a path instance $p \in \mathcal{P}$ from an object $a$ to another object $b$. 

Following this guideline, we propose applying such a meta-path based similarity framework to our application scenario and further measure the similarity of a question $q_x$ and another question $q_y$ under a meta path $QSQ$ in the problem-solving network (Figure \ref{fig1}). Each path instance $q_{x}s_{i}q_{y} \in QSQ$ passing a student $s_i$ is measured using the cosine similarity:
\begin{equation} \label{eq_sim2}
    f(q_{x}s_{i}q_{y})= \frac{\mathbf{Feature_{ix} }\cdot \mathbf{Feature_{iy} }}{\|\mathbf{Feature_{ix} }\|\|\mathbf{Feature_{iy}}\|}
\end{equation}
where $\mathbf{Feature_{ix}}$ is the feature vector, consisting of both mouse movement features generated based on the mouse trajectory of the submission on $q_x$ by a student $s_i$ (as introduced in Section 4.2) and the score of his/her submission.

We normalize the similarity score of each path instance $q_{x}s_{i}q_{y}$ by using the sum of the similarity scores of all the students who have finished both questions $q_{x},q_{y}$,
guaranteeing that every score in our problem-solving network is between 0 and 1. Thus, the final similarity between $q_x$ and $q_y$ on meta path $QSQ$ is defined as:
\begin{equation}
    s(q_{x}, q_{y})=\frac{\sum_{q_{x}s_{i}q_{y} \in QSQ} \frac{\mathbf{Feature_{ix} }\cdot \mathbf{Feature_{iy}}}{\|\mathbf{Feature_{ix} }\|\|\mathbf{Feature_{iy}}\|}}{\left|\{q_{x}s_{i}q_{y}: q_{x}s_{i}q_{y} \in QSQ \}\right|}
\end{equation}

% \yong{I revised the writing of the above, pls double check it and make sure I do not change the idea here.}

% \subsection{Prediction Algorithm}\label{algorithm}

% subsection{Feature Vector Concatenation}

%% file: src/05_evaluation.tex
\section{Experiments}
\subsection{Experiment Setup}
Our experiment was conducted on the two datasets introduced in Table~\ref{tabledataset}.
% and we managed to map the original score from 0 to 100 to 4 score classes (0-3) based on the existing score criteria on the platform. 
To make a 4-class classification, we applied four classical multi-class classification machine learning models, i.e., GBDT, RF, SVM, and LR, on our datasets. %, respectively. 
% \yong{Give a full name of these algorithms in their first appearance. Also, use one sentence to briefly introduce why these four algorithms? e.g., because they have been widely used in student performance prediction?}
First, we built question statistics and student statistics (Table~\ref{table_other_fea}) with all records before April 12, 2019. We then calculated the student recent statistics for each record after April 12, 2019, using the data in the past 14 days of that record. 
For those submissions without recent records, we assigned -1 to all recent performance features. The features listed in Table~\ref{table_other_fea} served as features in baseline method. 

\begin{table}[]
\small
\begin{tabular}{@{}cccccc@{}}
\toprule
Dataset & Method & GBDT & RF & SVM & LR \\ \midrule
\multirow{3}{*}{ADD} & Ours & 0.88 & 0.87 & 0.85 & 0.87 \\ \cmidrule(l){2-6} 
 & baseline & 0.79 & 0.85 & 0.77 & 0.83 \\ \cmidrule(l){2-6} 
 & ABROCA & \textbf{0.09} & \textbf{0.02} & \textbf{0.08} & \textbf{0.04} \\ \midrule
% \multirow{3}{*}{DGDD} & Ours & 0.86 & 0.88 & 0.90 & 0.92 \\ \cmidrule(l){2-6} 
%  & baseline & 0.88 & 0.88 & 0.89 & 0.91 \\ \cmidrule(l){2-6} 
%  & ABROCA & -0.02 & 0 & 0.01 & 0.01 \\ \bottomrule
 \multirow{3}{*}{DGDD} & Ours & 0.94 & 0.90 & 0.91 & 0.91 \\ \cmidrule(l){2-6} 
 & baseline & 0.88 & 0.88 & 0.89 & 0.89 \\ \cmidrule(l){2-6} 
 & ABROCA & \textbf{0.06} & \textbf{0.02} & \textbf{0.02} & \textbf{0.02} \\ \bottomrule
\end{tabular}
\caption{AUC and ABROCA value in two datasets. ADD: area dimension question dataset. DGDD: deductive geometry dimension question dataset. Ours: our proposed method. ABROCA: Area between baseline curve and Ours curve.}
\label{tableAUC}
\vspace{-3em}
\end{table}

Based on the proposed mouse movement features, similarity matrix $M_{sim}$ of each dataset was extracted according to the problem-solving network and related similarity calculation algorithms mentioned in Section 4.4. Then for each first submission $q_{x}-s_{i}$ in the dataset, we searched $M_{sim}$ to find question $q_y$ that was most similar to $q_x$ (with a similarity threshold of 0.8 in the experiment on ADD, 0.7 in the experiment on DGDD) and is solved by the same student $s_m$.  If there was no similar question with $q_x$ done by $s_{i}$, this submission record would be discarded. Finally, the feature vector of $q_{x}-s_{i}$ in our proposed method was composed of $q_{x}$'s and $s_{i}$'s historical statistical features, $s_{i}$'s recent performance feature in Table \ref{table_other_fea},  $q_y$'s  mouse movement features including features in Table \ref{table:3} and Table \ref{tableMmeasure}, $q_y$'s score class and the similarity score between $q_x$ and $q_y$.

Since we discarded submissions with no similar questions, the dataset of proposed method for training and testing is not of the same size as the original dataset. We used the same submission records in the baseline and the proposed methods. In addition, to reduce the effect of imbalanced class distribution, we applied SMOTE over-sampling algorithm to increase the size of minority classes in the training set \cite{smote}.
As for hyperparameter tuning in the four algorithms, grid-search was conducted in model training on our datasets~\cite{manrique2019analysis}. The following parameters have been tested with the best values in bold:
\begin{itemize}
\item Number of trees for GBDT and RF: 50, 100, 150, 200, \textbf{250}, 300, 350
\item Max depth of trees for GBDT and RF: \textbf{5}, 10, 15, 20, 25
\item Learning rate of GBDT: 1e-4, \textbf{1e-3}, 1e-2, 5e-2, 0.1, 0.2
\item Penalty parameter of SVM (C): 0.1, 1, \textbf{5}, 10
\end{itemize}

After fixing our hyperparameters, each model performed the task of predicting students' performance on two datasets using the baseline method and our proposed method for 10 times each and the average accuracy scores and weighted F1 scores are in Table \ref{table_acc}.
\subsection{Performance Comparison}
The results for our dataset ADD and DGDD are in Table \ref{table_acc}. From the perspectives of accuracy and weighted F1 scores, we can see that in both two datasets, GBDT, RF, and SVM performed better in our proposed method than in the baseline. In addition, the performances of our method and the baseline are similar in both datasets.
\begin{table*}[]
\small
\begin{tabular}{@{}cccccccccc@{}}
\toprule
\multirow{2}{*}{Dataset} & \multirow{2}{*}{Method} & \multicolumn{2}{c}{GBDT} & \multicolumn{2}{c}{RF} & \multicolumn{2}{c}{SVM} & \multicolumn{2}{c}{LR} \\ \cmidrule(l){3-10} 
 &  & Accuracy & Weighted F1 & Accuracy & Weighted F1 & Accuracy & Weighted F1 & Accuracy & Weighted F1 \\ \midrule
\multirow{2}{*}{ADD} & Baseline & 0.555 & 0.555 & 0.669 & 0.659 & 0.430 & 0.492 & \textbf{0.650} & \textbf{0.670} \\ \cmidrule(l){2-10} 
 & Ours & \textbf{0.753} & \textbf{0.749} & \textbf{0.690} & \textbf{0.667} & \textbf{0.650} & \textbf{0.677} & 0.649 & 0.659 \\ \midrule
\multirow{2}{*}{DGDD} & Baseline & 0.600 & 0.597 & 0.664 & 0.663 & 0.600 & 0.611 & 0.720 & 0.731 \\ \cmidrule(l){2-10} 
 & Ours & \textbf{0.833} & \textbf{0.805} & \textbf{0.780} & \textbf{0.767} & \textbf{0.633} & \textbf{0.643} & \textbf{0.733} & \textbf{0.744} \\ \bottomrule
% \multirow{2}{*}{DGDD} & Baseline & 0.672 & 0.675 & 0.652 & 0.651 & 0.560 & 0.567 & \textbf{0.720} & \textbf{0.731} \\ \cmidrule(l){2-10} 
%  & Ours & \textbf{0.720} & \textbf{0.698} & \textbf{0.720} & \textbf{0.721} & \textbf{0.720} & \textbf{0.721} & 0.680 & 0.693 \\ \bottomrule
\end{tabular}
\caption{Results of the accuracy and weighted F1 over four typical machine learning algorithms (GBDT, RF, SVM, and LR) on the proposed method and the baseline method. ADD: area dimension question dataset. DGDD: deductive geometry dimension question dataset. Ours: our proposed method.}
\label{table_acc}
\end{table*}

\begin{figure}[ht]
\centering
\includegraphics[width=0.8\linewidth]{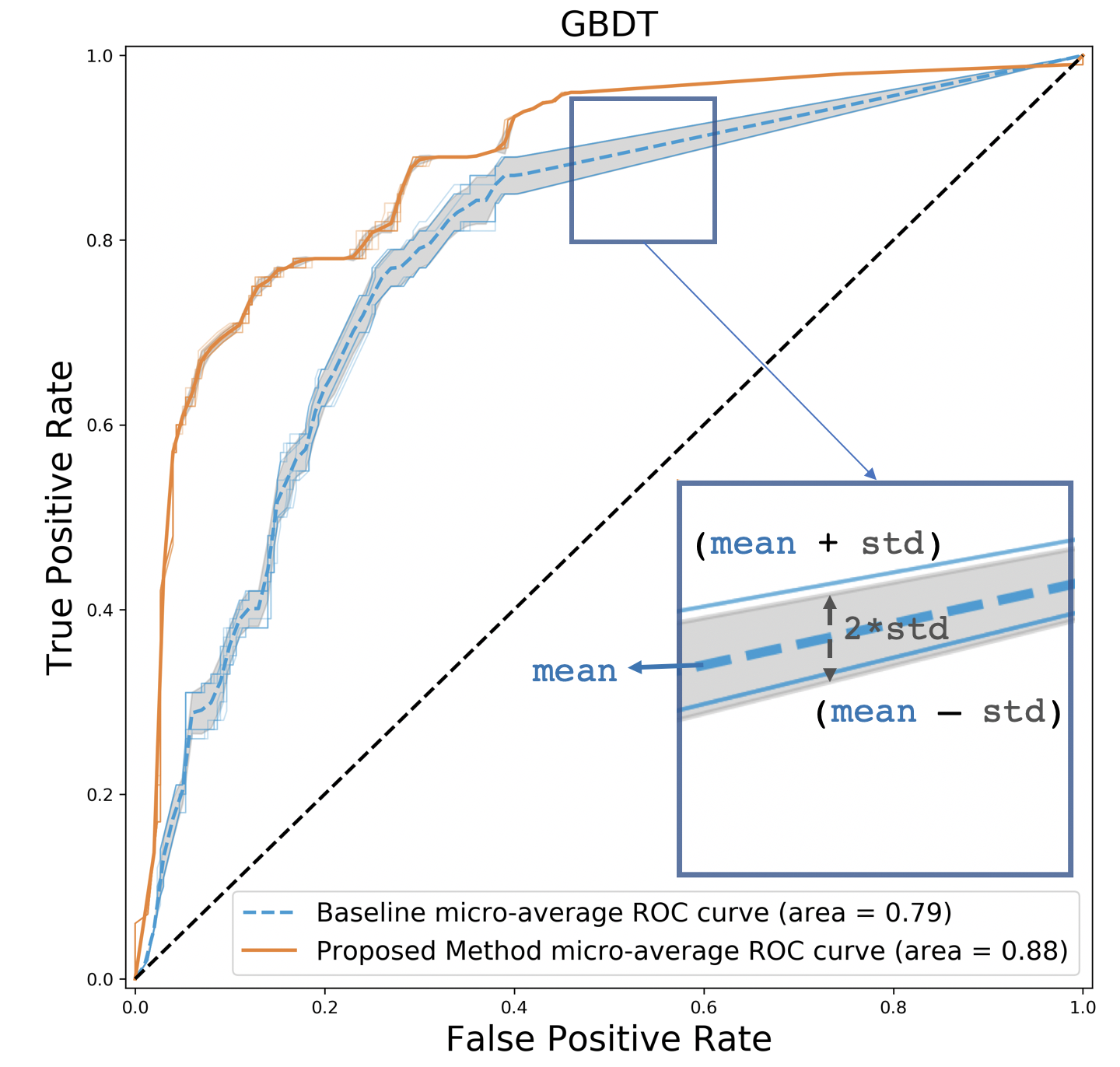}
\vspace{-1em}
\caption{ROC-AUC curve of two methods on ADD (Area dimension question dataset).Gray borders represent mean+std or mean-std. Area between the two curves is ABROCA.}
\label{ROC}
\vspace{-2em}
\end{figure}
\begin{figure*}
  \begin{subfigure}[b]{0.45\textwidth} \includegraphics[width=\textwidth]{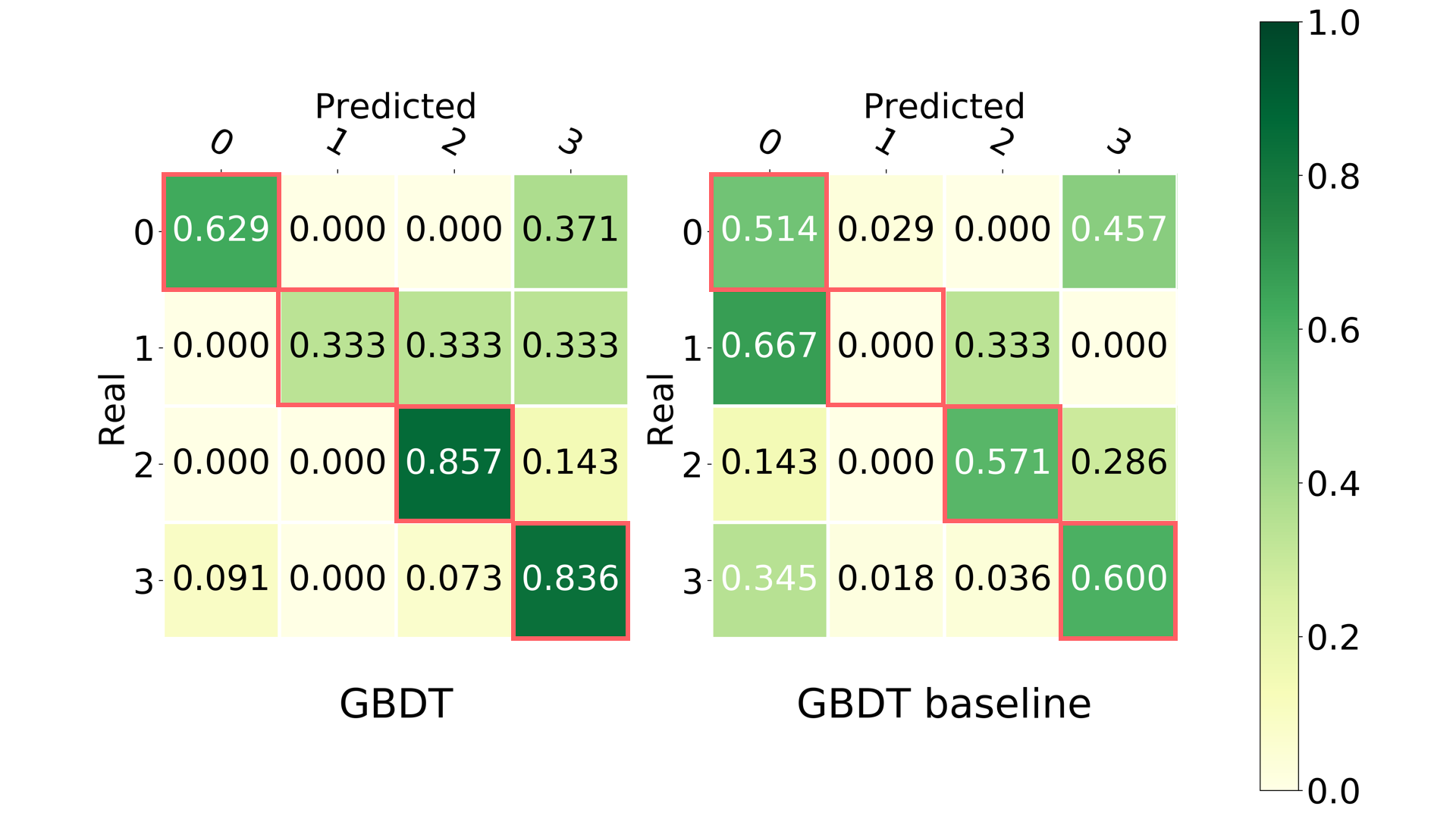}
    \vspace{-3em}
    \caption{GBDT}
    \label{heatmap_GBDT}
  \end{subfigure}
  \begin{subfigure}[b]{0.45\textwidth}
    \includegraphics[width=\textwidth]{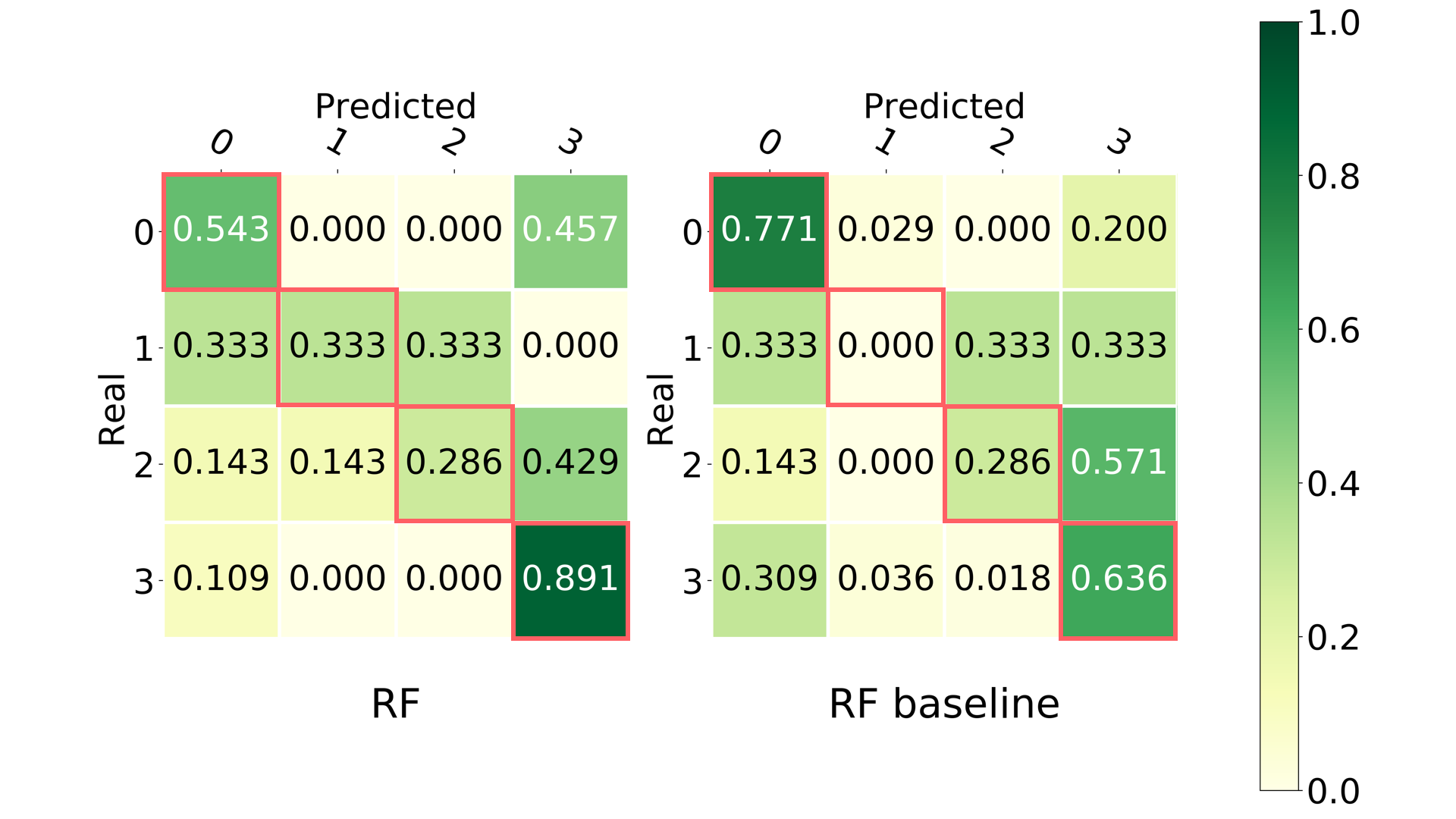}
    \vspace{-3em}
    \caption{RF}
    \label{heatmap_RF}
  \end{subfigure}

  \begin{subfigure}[b]{0.45\textwidth}
    \includegraphics[width=\textwidth]{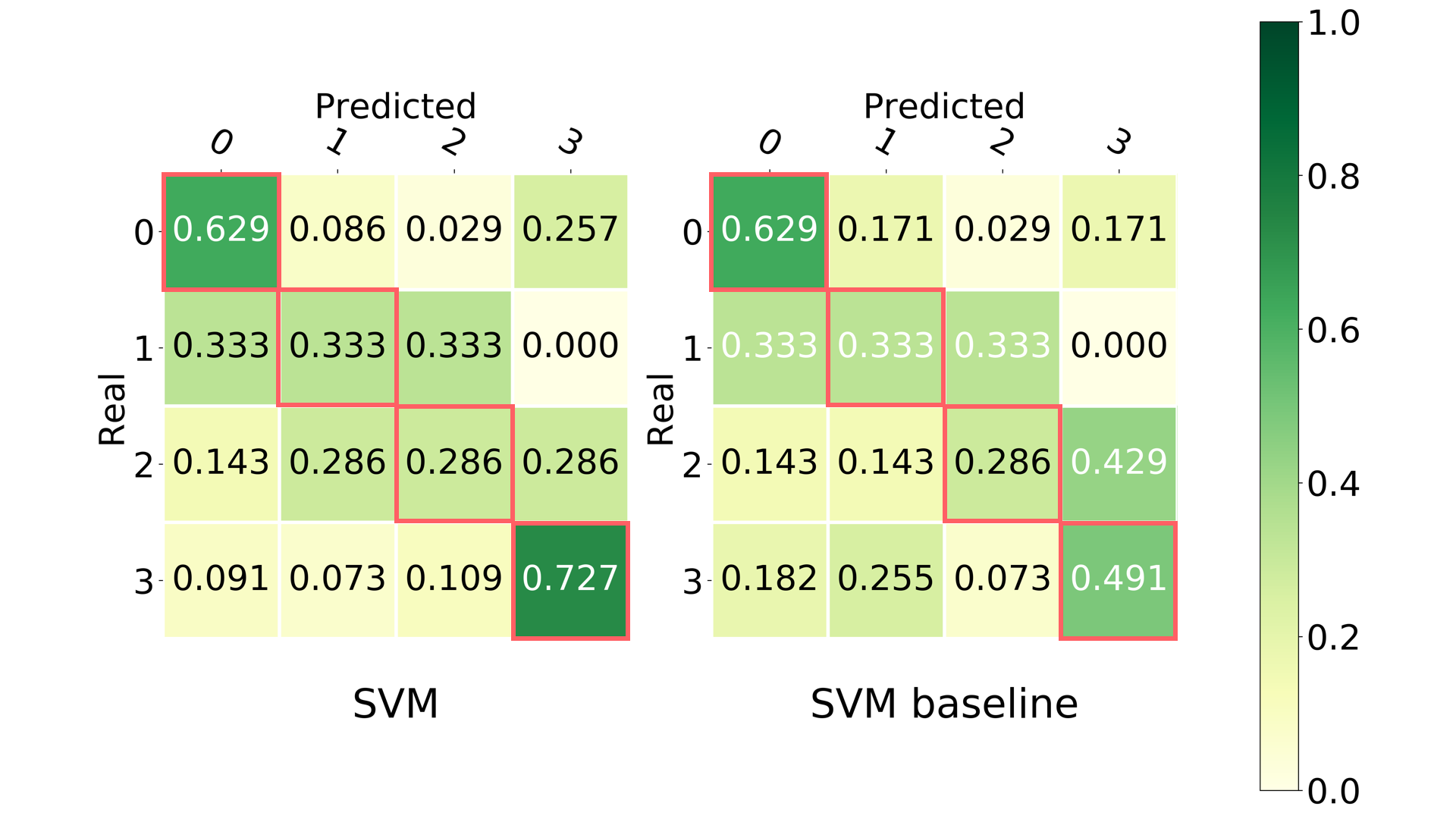}
    \vspace{-3em}
    \caption{SVM}
    \label{heatmap_SVM}
  \end{subfigure}
  \begin{subfigure}[b]{0.45\textwidth}
    \includegraphics[width=\textwidth]{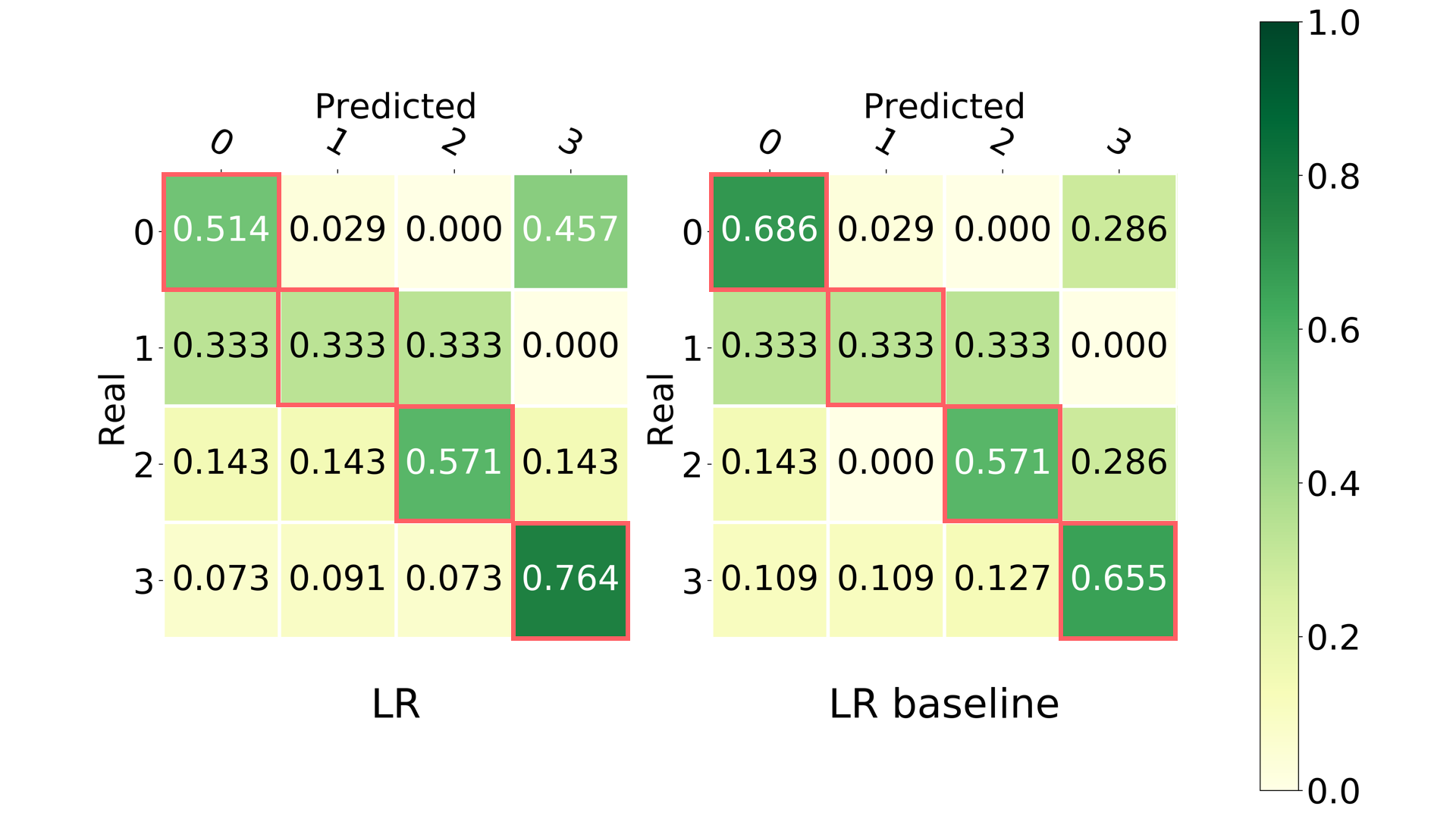}
    \vspace{-3em}
    \caption{LR}
    \label{heatmap_LR}
  \end{subfigure}
  \vspace{-1em}
  \caption{Predicted distribution heatmap of each score class with four algorithms on dataset ADD. The proportion of True Positive samples in each score class is in red box. In (a)-(d), the heatmap on the left shows the result of our proposed method and the right one shows the baseline method. ADD: area dimension question dataset. 
  %ADD test set real distribution: \{0:35, 1:3, 2: 7, 3:55\}
  }
  \label{heatmap}
\end{figure*}

Besides the overall accuracy, we further evaluate the results based on the Receiver Operating Characteristic (ROC) curve. The ROC curve is a graph showing the performance of a classification model at all classification thresholds.
%This curve plots two parameters, false positive rate and true positive rate. 
The ROC curve is a plot of the false positive rate and true positive rate. 
The area under the ROC curve (AUC) measures the entire two-dimensional area underneath the entire ROC curve from $(0,0)$ to $(1,1)$ \cite{rocauc,gardner2019evaluating}, which means AUC ranges from 0 to 1. AUC provides an aggregate measure of performance across all possible classification, so we use the area between the proposed method's ROC curve and the baseline ROC curve to further compare the performance of our proposed method with baseline's, which is called ABROCA in prior work~\cite{gardner2019evaluating}. To eliminate the randomness of the algorithms, we ran the program 10 times and made a $mean+std$ and $mean-std$ ROC-AUC graph (Figure \ref{ROC}). As Table \ref{tableAUC} shows, the ABROCA value is always positive, this confirms that the aggregate performance of our proposed method is consistently better than the baseline across different models.
Furthermore, we extended the student score prediction from a binary classification problem (correct or wrong) to a multiple classification problem (0-3). To further evaluate our method's performance in every score class, we selected ADD (area dimension question dataset) and drew a real-predicted heatmap (Figure \ref{heatmap}) for each algorithm.
% Specially, when we go deeper into the score prediction distribution performance of the four algorithms, 
% work well across different models.
% \begin{table}[]
% \begin{tabular}{|c|c|c|c|c|c|}
% \hline
% Dataset& Method & GBDT & RF & SVM & LR \\ \hline
% \multicolumn{1}{|c|}{\multirow{3}{*}{ADD}} & Ours & 0.88 & 0.87 & 0.85 & 0.87 \\ \cline{2-6} 
% \multicolumn{1}{|c|}{} & baseline & 0.79 & 0.85 & 0.77 & 0.83 \\ \cline{2-6} 
% \multicolumn{1}{|c|}{} & ABROCA & 0.09 & 0.02 & 0.08 & 0.04 \\ \hline
% \multicolumn{1}{|c|}{\multirow{3}{*}{DGDD}}& Ours & 0.86 &  0.88&  0.90& 0.92 \\ \cline{2-6} 
% \multicolumn{1}{|c|}{} & baseline &  0.88& 0.88 & 0.89 & 0.91 \\ \cline{2-6}
% \multicolumn{1}{|c|}{} & ABROCA &  &  &  &  \\ \hline
% \end{tabular}
% \caption{AUC and ABROCA value in two dataset. ADD: area dimension question dataset. DGDD: deductive geometry dimension question dataset. Ours: our proposed method. ABROCA: Area between baseline and proposed method.}
% \label{table:1}
% \end{table}
% \usepackage{booktabs}
% \usepackage{multirow}
% \usepackage{booktabs}
% \usepackage{multirow}
% \subsection{Feature Analysis}
% Specially, we learned that GBDT can provide feature
Specially, the GBDT model provides an importance score for each feature, which is computed by the normalized summation of reduction in loss function of each feature and it is also called Gini importance\cite{scikit-learn}. The importance score of each feature ranges from 0 to 1 and a larger importance score indicates that the corresponding feature is more important in training the model. This directly helps to learn the importance distribution of our feature sets. In our method, the importance score of mouse movement feature set, most similar question's score class and the similarity score (39 features) reaches 27.4\% among 483 features, which indicates the mouse movement feature set has significant contribution in GBDT model.
\vspace{-1em}
% It directly helps us 
% we also calculate all features' importance in the GBDT model on dataset ADD. 
% Feature importance in the model is computed by the normalized summation of reduction in loss function of each feature, which is also called Gini importance\cite{scikit-learn}. 
% 39,13,6,418, Among 483 features,    %From the tree map(figure ~\ref{feature_importance}), we can see that 
%1. Compare different feature set.
%2. Talk why our feature is important
\iffalse
\begin{figure}[h]
\centering
\includegraphics[width=0.8\linewidth]{src/image/treemap.png}
\caption{The treemap of feature importance in algorithm GBDT on the question set Area-dimension. The color encoding of each area:  yellow ( question statistics), green (mouse movement), red (student statistics), blue (student recent statistics).  Top 4 important features are labeled in the graph as (1)-(4): (1) Proportion of submissions in score class 1 (2) Score class of the most similar question, (3) Proportion of submissions in score class 2, (4) Proportion of submissions in numeric questions of grade 2 and difficulty 5}
\label{feature_importance}
\end{figure}
\fi

%% file: src/06_discussion.tex
\section{Discussion}
The above experiment results demonstrate that our approach can achieve higher accuracy for predicting student performance prediction in interactive online question pools than using the baseline features. However, there are still some issues that need further discussions.

% \textbf{Interaction Feature Design}
% 或者叫做
% \textbf{Think Time vs. Response Time} \yong{Huan, please discuss what you want to talk here.}

\textbf{Parameter Configurations} Some parameters that need to be set in the proposed approach and some of these parameters are empirically chosen after considering different factors. For example, the sliding window is used to detect the change points (Section~\ref{sec_change_point}), where the value of the window size and threshold need to be determined first.
Too large a window size will make the mouse drag event histogram too smooth, while too small a window size may make the mouse drag event frequency histogram too steep. Both have a negative effect on change point detection. 
The corresponding threshold is also important. We empirically set it as the average mouse drag density to guarantee that the detected change points are exactly the actual ones.
With similar considerations, we choose only the question with the highest similarity score and require that the similarity score should be at least $0.7$, 
when we try to incorporate the information of similar questions.
Our experiment results provide support for the effectiveness of the current parameter settings. 
% A more adaptive parameter setting is further exploration.

\textbf{Prediction Models} This paper focuses on proposing novel methods to extract features for student performance prediction. These features are further combined with existing well-established traditional machine-learning based prediction models to predict student future performance. Compared with other methods based on deep neural networks (e.g., deep knowledge tracing~\cite{piech2015deep,yeung2018addressing}), we argue that predictions based on feature extraction have better explainability, as the features are often intuitive and meaningful.

% \textbf{Evaluation Design} 

\textbf{Data Issues} Since there are no other dataset of interactive online question pools publicly available, the whole study is conducted on the data collected from only one interactive online math question pool. But our proposed approach can be easily extended to other datasets of interactive online question pools, which mainly involve drag-and-drop interactions.
Moreover, the interaction features based on student mouse movement trajectories rely on the size of the interaction records. When there are too few student interaction records for a specific question, it will be difficult for us to accurately compare the similarity between a certain question and others using student interaction features. With more student interaction data being collected, the reliability of the proposed approach can be further improved. In addition, our method can be easily extended to other devices such as tablets with touch screens since the collected data has the same format (i.e., timestamp, event, and position).
\vspace{-1em}

%% file: src/07_conclusion.tex
\section{Conclusion}

Different from the extensively-studied student performance prediction in MOOCs platforms, student performance prediction in interactive online question pools can be more challenging due to the lack of knowledge tags and predefined question order or course curriculum. 
We proposed a novel method to boost student performance prediction in interactive online question pools by incorporating student interaction features and similarity between questions. 
We extracted new features based on student mouse movement trajectories to delineate problem-solving details of students. We also applied HIN to further consider students' historical problem-solving information on similar questions, as students' recent performance on similar questions can also be a good indicator of student future performance on a certain question.
We conducted extensive experiments with the proposed method on the dataset collected from a real-world interactive online math question pool.
Compared with using only the traditional statistic features (e.g., average scores), the proposed method achieved a much higher prediction accuracy across different models in different question classes. The results further confirm the effectiveness of our method.

In future work, we would like to combine the proposed method with adaptive question recommendation in interactive online question pools, providing different students with personalized online learning and question practice in online question pools. Also, it would be interesting to summarize different problem-solving patterns using mouse trajectories to better model and understand students' learning behaviors.

% In this paper, we propose a novel mouse movement based feature sets for predicting student performance. Specifically, we introduce a wait-time based feature and ......